\input{aipcheck}

\documentclass[
    ,final            
  ]
  {aipproc}

\layoutstyle{6x9}

\begin{document}

\title{Korean-Japanese Planet Search Program: Substellar Companions around Intermediate-Mass Giants}


\author{Masashi Omiya}{
  address={Korea Astronomy and Space Science Institute, 61-1 Hwaam-dong, Yuseong-gu, Daejeon 305-348, South Korea, email: {omiya@kasi.re.kr}}
}

\author{Inwoo Han}{
  address={Korea Astronomy and Space Science Institute, 61-1 Hwaam-dong, Yuseong-gu, Daejeon 305-348, South Korea, email: {omiya@kasi.re.kr}}
}

\author{Hideyuki Izumiura}{
  address={Okayama Astrophysical Observatory, National Astronomical Observatory of Japan, Asakuchi, Okayama 719-0232, Japan}
  ,altaddress={Department of Astronomical Science, The Graduate University for Advanced Studies, Shonan Village, Hayama, Kanagawa 240-0193, Japan} 
}

\author{Byeong-Cheol Lee}{
  address={Korea Astronomy and Space Science Institute, 61-1 Hwaam-dong, Yuseong-gu, Daejeon 305-348, South Korea, email: {omiya@kasi.re.kr}}
}
\author{Bun'ei Sato}{
  address={Tokyo Institute of Technology, 2-12-1-S6-6 Ookayama, Meguro-ku, Tokyo 152-8550, Japan}
}
\author{Kang-Min Kim}{
  address={Korea Astronomy and Space Science Institute, 61-1 Hwaam-dong, Yuseong-gu, Daejeon 305-348, South Korea, email: {omiya@kasi.re.kr}}
}
\author{Tae Seog Yoon}{
  address={Department of Astronomy and Atmospheric Sciences, Kyungpook National University, Daegu 702-701, South Korea}
}

\author{Eiji Kambe}{
  address={Okayama Astrophysical Observatory, National Astronomical Observatory of Japan, Asakuchi, Okayama 719-0232, Japan}
}
\author{Michitoshi Yoshida}{
  address={Hiroshima Astrophysical Science Center, Hiroshima University, Higashi-Hiroshima, Hiroshima 739-8526, Japan}
}
\author{Seiji Masuda}{
  address={Tokushima Science Museum, Asutamu Land Tokushima, Itano-gun, Tokushima 779-0111, Japan}
}
\author{Eri Toyota}{
  address={Kobe Science Museum, 7-7-6 Minatojimanakamachi, Chuo-ku, Kobe, Hyogo 650-0046, Japan}
}
\author{Seitaro Urakawa}{
  address={Bisei Spaceguard Center, Japan Spaceguard Association, 1716-3 Okura, Bisei-cho, Ibara, Okayama 714-1411, Japan}
}
\author{Masahide Takada-Hidai}{
  address={Liberal Arts Education Center, Tokai University, 1117 Kitakaname, Hiratsuka, Kanagawa 259-1292, Japan}
}

\begin{abstract}
A Korean-Japanese planet search program has been carried out using the 1.8m telescope at Bohyunsan Optical Astronomy Observatory (BOAO) in Korea, and the 1.88m telescope at Okayama Astrophysical Observatory (OAO) in Japan to search for planets around intermediate-mass giant stars. The program aims to show the properties of planetary systems around such stars by precise Doppler survey of about 190 G or K type giants together with collaborative surveys of the East-Asian Planet Search Network. So far, we detected two substellar companions around massive intermediate-mass giants in the Korean-Japanese planet search program. One is a brown dwarf-mass companion with 37.6 $M_{\mathrm{J}}$ orbiting a giant HD 119445 with 3.9 $M_{\odot}$, which is the most massive brown dwarf companion among those found around intermediate-mass giants. The other is a planetary companion with 1.8 $M_{\mathrm{J}}$ orbiting a giant star with 2.4 $M_{\odot}$, which is the lowest-mass planetary companion among those detected around giant stars with $>$ 1.9 $M_{\odot}$. Plotting these systems on companion mass vs. stellar mass diagram, there seem to exist two unpopulated regions of substellar companions around giants with 1.5--3 $M_{\odot}$ and planetary companions orbiting giants with 2.4--4 $M_{\odot}$. The existence of these possible unpopulated regions supports a current characteristic view that more massive substellar companions tend to exist around more massive stars.
\end{abstract}

\maketitle


\section{Introduction}
Many of the over 380 exoplanets discovered to date orbit solar mass (0.7--1.5 $M_{\odot}$) stars. This observational bias occurs because the main targets of previous Doppler spectroscopy-based exoplanet searches have been solar-type stars. These studies have revealed a variety of their planetary systems (e.g. \citealt{Butler2006}; \citealt{Udry2007}) whose statistical properties are now used to constrain planet-formation models (e.g. \citealt{Ida2004}). However, only about 35 and 25 planets were detected around intermediate-mass (1.5--5 $M_{\odot}$) and low-mass ($<$ 0.7 $M_{\odot}$) stars in surveys of evolved G-K type (sub) giants and K-M type dwarfs, respectively, so far. The planetary systems around such stars are not well understood. Clarifying the relationship between stellar mass and planetary system is important for understanding general planet formation because planet formation should depend on the properties of the protoplanetary disks, which be affected by the condition of host stars.

Evolved intermediate-mass (sub) giant stars are suitable targets for Doppler spectroscopy-based planet searches because these stars have low surface activity and their spectra exhibit many sharp absorption lines. Although the number of substellar companions found orbiting such stars is still small, some planetary system properties have begun to emerge. For example, the masses of planets and their host stars are correlated: more massive substellar companions tend to exist around more massive stars (e.g. \citealt{Lovis2007}). This correlation suggests that the mass ranges of the brown dwarf desert depend on the host star mass \citep{Omiya2009}. The planet occurrence rate also depends on the host star mass: the planet frequency around 1.5--2 $M_{\odot}$ giant stars is higher than that around lower-mass stars \citep{Johnson2010a}. Moreover, the semi-major axes of planetary systems seem to be correlated with host star: the range for planets orbiting intermediate-mass giant stars exceeds 0.6 AU\footnote[1]{Recently, a close-in planet with a semi-major axis of 0.081 AU was found orbiting an intermediate-mass subgiant with mass of 1.68 $M_{\odot}$ \citep{Johnson2010b}.}, and the range for planets around solar-type stars is large than 0.02 AU (e.g. \citealt{Sato2008}). These properties of substellar systems around intermediate-mass giant stars seem to be not similar to those around solar-type stars. A unified understanding of planetary systems over a wide range of host star masses will provide valuable insights into the dependence of substellar systems on the central stars and into general plant formation mechanisms.

\section{Korean-Japanese Planet Search Program}
In 2005, we started a joint planet search program between Korean and Japanese researchers to search for planets around GK-type giant stars using a precise Doppler technique with using the 1.8-m telescope at BOAO and the 1.88-m telescope at OAO \citep{Omiya2009}. This survey program is an extended version of the ongoing OAO planet search program \citep{Sato2005} and part of an international collaboration among researchers from Korea, China and Japan (an East-Asian Planet Search Network, EAPS-Net; \citealt{Izumiura2005}). The collaboration aims at clarifying the properties of planetary systems around intermediate-mass stars by surveying more than 800 GK giants for planets at OAO, BOAO, the Xinglong station (China), and the Subaru Telescope.

For the Korean-Japanese planet search program, we selected about 190 target stars from the $Hipparcos$ catalog based on the following criteria: color-index 0.6 $<$ $B-V$ $<$ 1.0, absolute magnitude $-$3 $<$ $M_{v}$ $<$ 2, declination $\delta$ $>$ $-$25 $^{\circ}$, and visual magnitude 6.2 $<$ $V$ $<$ 6.5 \citep{Omiya2009}. These targets are fainter than those of the OAO and Xinglong program (\citealt{Sato2005}; \citealt{Liu2008}). We divided the targets into two parts: one for BOAO and the other for OAO. Each is observed independently at the assigned observatory, although a star that exhibits a large radial velocity variation is observed intensively at both observatories.

\subsection{BOES Observations and Analysis}
Radial velocity observations at BOAO are carried out with the 1.8-m telescope and the BOAO Echelle Spectrograph (BOES; \citealt{Kim2007}), a fiber-fed high resolution echelle spectrograph. We place an iodine (I$_{2}$) cell in the optical path in front of the fiber entrance of the spectrograph \citep{Kim2002} for precise wavelength calibration and use a 200-$\mu$m fiber, obtaining a wavelength resolution $R$ = $\lambda$/$\Delta{\lambda}$ $\sim$ 51000. The spectra cover the 3500--10500 \AA \ wavelength range. Echelle data reduction is performed using the IRAF software package in the standard manner. We use the 5000 \AA \ to 5900 \AA \ wavelength range, a region covered by many I$_{2}$ absorption lines, for radial velocity measurements. We also make use of Ca II H lines at around 3970 \AA \ as chromospheric activity diagnostics. Radial velocity analysis is performed using the spectral modeling technique described in \citet{Sato2002}, which is based on the method of \citet{Butler1996}, and improve and optimize for BOES data analysis \citep{Omiya2009}. We employ the extraction method described in \citet{Sato2002} to prepare a stellar template spectrum from stellar spectra taken through the I$_{2}$ cell (star+I$_{2}$ spectra). The technique allows us to achieve a long-term Doppler precision of 14 m s$^{-1}$ over 5.5 yr.

\subsection{HIDES Observations and Analysis}
Radial velocity observations at OAO are carried out with the 1.88-m telescope and HIgh Dispersion Echelle Spectrograph (HIDES; \citealt{Izumiura1999}) attached to the Coud\'e focus of the telescope. We use an I$_{2}$ cell placed in the optical path in front of the slit of the spectrograph \citep{Kambe2002} as a precise wavelength calibrator. We set the slit width to 200 $\mu$m (0.76"), providing a spectral resolution of 63000. Until November 2007, we had taken star+I$_{2}$ spectra with the 5000 \AA \ to 6200 \AA \ wavelength range, and had taken stellar spectra at the same wavelength range without the I$_{2}$ cell for abundance analysis. Since the HIDES CCD system was upgraded to a three CCD mosaic in December 2007, we obtain simultaneous spectra in the full 3750 \AA \ to 7550 \AA \ wavelength range. We use the 5000 \AA \ to 5900 \AA \ wavelength ranges of star+I$_{2}$ spectra for radial velocity measurements. Echelle data reduction is performed using the IRAF software package in the standard manner. Stellar radial velocities are derived from the I$_{2}$-superposed stellar spectrum using the spectral modeling techniques detailed in \citet{Sato2002}, giving a Doppler precision of less than 8 m s$^{-1}$ over 5.5 yr.

\section{Two Substellar Companions around Massive Intermediate-Mass Giants}
Up to now, we have been monitoring the radial velocities of sample stars using 1.8-m BOAO telescope and 1.88-m OAO telescope for 5.5 yr and identified many candidate stars with large radial velocity variations. And then, two massive intermediate-mass stars with a periodic variation caused by a substellar companion are found among them: a brown dwarf-mass companion orbiting HD119445 \citep{Omiya2009} and a planetary companion orbiting a giant star.

\begin{figure}
  \includegraphics[width=.8\textwidth]{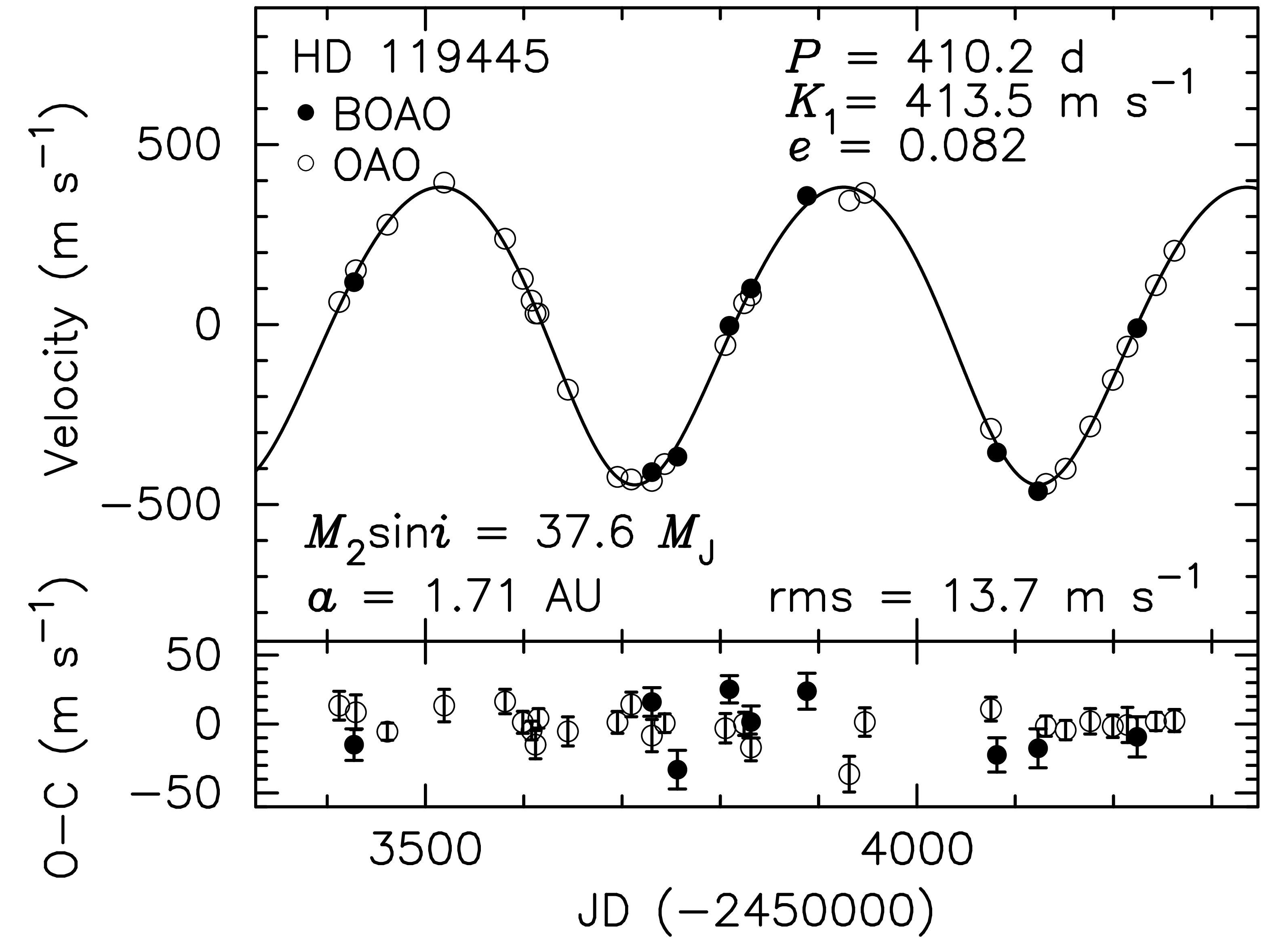}
  \caption{Upper panel: radial velocities of HD 119445 observed at BOAO (\textit{filled circles}) and OAO (\textit{open circles}). The solid line represents the Keplerian orbital curve. Lower panel: Residuals to the best Keplerian fit.}
  \label{fig1}
\end{figure}

We monitored the radial velocity of HD 119445 for 2.3 years from the beginning of the survey at both observatories. The observed radial velocities of HD 119445 are shown in figure \ref{fig1}. The best-fit Keplerian orbit derived from both the BOAO and OAO data has a period $P$ $=$ 410.2 $\pm$ 0.6 days, a velocity semi-amplitude $K_{1}$ $=$ 413.5 $\pm$ 2.6 m s$^{-1}$, and an eccentricity $e$ $=$ 0.082 $\pm$ 0.007. The best-fit curve is shown in figure \ref{fig1} as a solid line overlaid on the observed velocities. The rms scatter of the residuals to the best-fit is 13.7 m s$^{-1}$. This values is comparable to the typical radial velocity scatter (10--20 m s$^{-1}$) of the G-type giants \citep{Sato2005}. Adopting a stellar mass $M$ = 3.9 $\pm$ 0.4 $M_{\odot}$ for HD 119445, we obtained a semi-major axis $a$ = 1.71 $\pm$ 0.06 AU and a minimum mass $M_{\mathrm{2}} \mathrm{sin} i$ = 37.6 $\pm$ 2.6 $M_{\mathrm{J}}$ for a brown dwarf-mass companion \citep{Omiya2009}.

\begin{figure}
  \includegraphics[width=.8\textwidth]{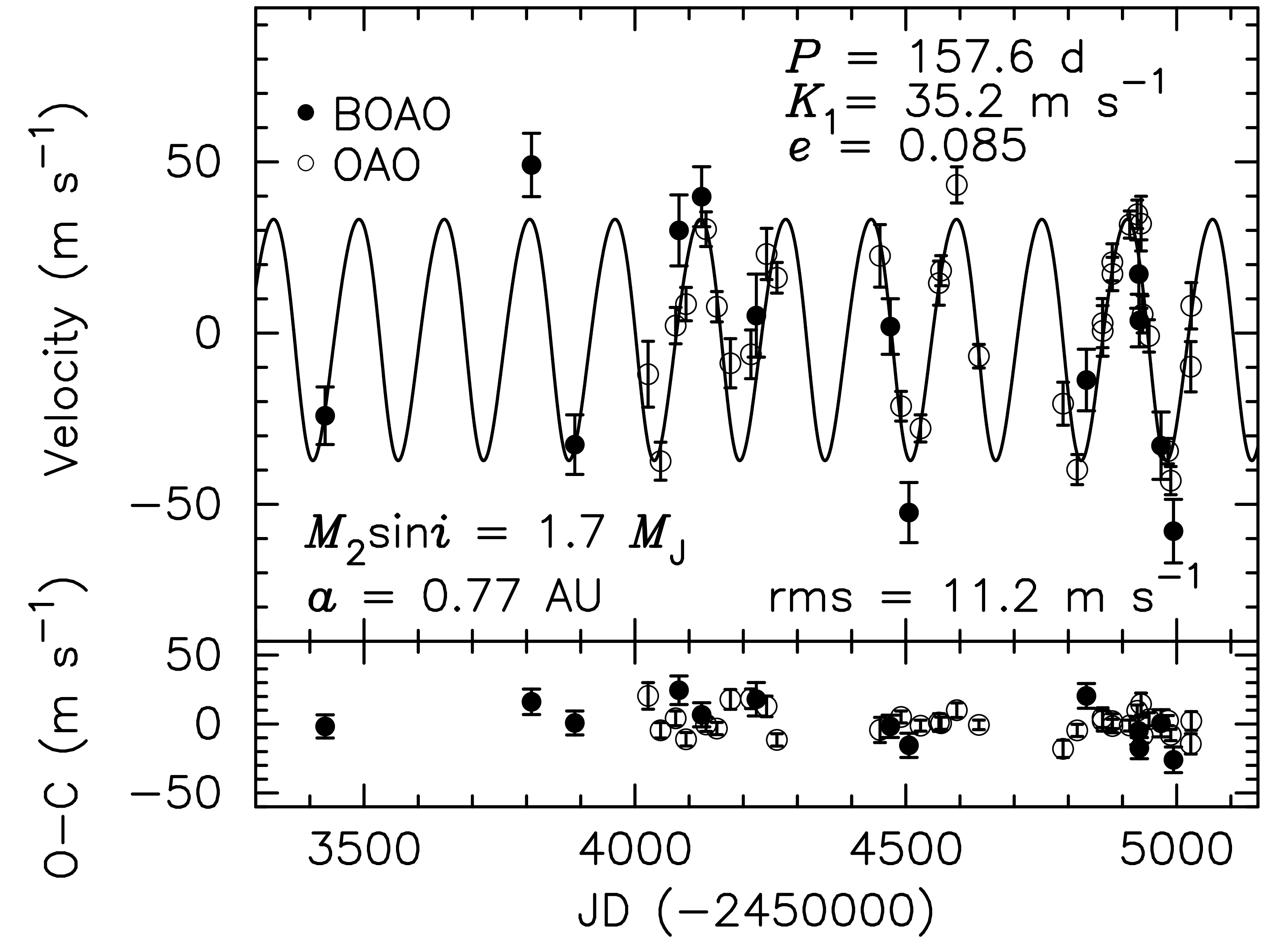}
  \caption{Upper panel: radial velocities of a giant star observed at BOAO (\textit{filled circles}) and OAO (\textit{open circles}). The solid line represents the Keplerian orbital curve. Lower panel: Residuals to the best Keplerian fit.}
  \label{fig2}
\end{figure}

A large radial velocity variation in the giant star with a planetary companion was found in the early BOAO radial velocity survey and we made intensive follow-up observations of the star at BOAO and OAO. The observed radial velocities of the star are shown in figure \ref{fig2}. A best-fit Keplerian orbit derived from both the BOAO and OAO velocity data by a least square fit has a period $P$ $=$ 157.57 d, a velocity semi-amplitude $K_{1}$ $=$ 35.2 m s$^{-1}$, and an eccentricity $e$ $=$ 0.085. The best-fit curve is shown in figure \ref{fig2} as a solid line overlaid on the observed velocities. The rms of the residuals to the best-fit are 11.2 m s$^{-1}$ for BOAO and OAO data. This values is comparable to the typical radial velocity scatter of the G-type giants \citep{Sato2005}. Adopting a host star's mass $M$ = 2.4 (2.0--2.6) $M_{\odot}$, which was estimated from evolutionary track and fundamental stellar parameters of $L$ = 43 $L_{\odot}$, $T_{\textrm{eff}}$ = 4861 K, and [Fe/H] = 0.15, we obtained a semi-major axis $a$ = 0.77 AU and a minimum mass $M_{\mathrm{2}} \mathrm{sin} i$ = 1.7 $M_{\mathrm{J}}$ for a planetary companion.

\section{Discussion}
Their host stars, HD119445 and the giant star, have masses of 3.9 $M_{\odot}$ and 2.4 $M_{\odot}$, respectively. The brown dwarf-mass companion orbiting HD 119445 is most massive substellar companions, and the planetary companion orbiting the giant star is lowest-mass companion, among those discovered around massive intermediate-mass (1.9--5 $M_{\odot}$) stars.
In figure \ref{fig1}, we plot masses of the companions detected within semi-major axis of 3 AU by precise Doppler surveys against their host star's masses; solar-mass stars (0.7 $M_{\odot}$ $\leq$ $M$ $<$ 1.5 $M_{\odot}$, \textit{open triangles}), intermediate-mass subgiants and giants (1.5 $M_{\odot}$ $\leq$ $M$ $\leq$ 5 $M_{\odot}$, \textit{filled circles}), intermediate-mass dwarfs (\textit{open circles}), and HD 119445 and the giant star (\textit{stars}) (e.g. \textit{The Extrasolar Planets Encyclopadia}; updated version of figure 5 in \citealt{Omiya2009}; this work). The detectable companion mass for a given host star mass depends on the orbital separation of its companion and the radial velocity jitter of the host star. Assuming that typical radial velocity jitters $\sigma$ for solar-mass stars, intermediate-mass subgiants (1.5--1.9 $M_{\odot}$) and giants (1.9--5 $M_{\odot}$) are $\sim$ 5 m s$^{-1}$, $\sim$ 7 m s$^{-1}$ and $\sim$ 20 m s$^{-1}$, respectively, we estimated the lower limits of companion masses detectable by precise Doppler surveys around a solar-mass star and intermediate-mass subgiant and giant at 3 AU (sold lines in figure \ref{fig1}), corresponding to companion masses that provide the amplitude of three times of typical radial velocity jitters. We also indicate detectable masses for these stars at 0.02 AU (\textit{dotted lines}) and 0.6 AU (\textit{dot-dashed lines}), corresponding to the semi-major axes of the known innermost planets orbiting solar-type and intermediate-mass evolved stars.

Two unpopulated regions of substellar companions orbiting intermediate-mass subgiants and giants seem to exist in region (a) and (b)\footnote[2]{We exclude a brown dwarf-mass companion orbiting a possible high mass giant HD 13189 ($M$ $=$ 4.5 $\pm$ 2.5 $M_{\odot}$; \citealt{Hatzes2005}) from the following discussion because of the large uncertainty in its host star's mass.}. A possible host star-companion mass correlation considered from unpopulated region (a) and (b) supports the current view that more massive substellar companions tend to exist around more massive stars, that are derived from the results of planet searches around various mass stars (\citealt{Lovis2007}; \citealt{Hekker2008}). 

All of the brown dwarf-mass companions to intermediate-mass evolved stars were found around those with $\geq$ 2.7 $M_{\odot}$ and there seems to be a paucity of such companions around those with 1.5--2.7 $M_{\odot}$ (region (a) in figure \ref{fig1}). Considering the smaller number of survey targets of $\geq$ 2.7 $M_{\odot}$ (e.g., 35 \% of the 300 OAO targets; \citealt{Takeda2008}) compared with that of 1.5--2.7 $M_{\odot}$, frequency of brown dwarf companions may become higher as stellar mass increases. This might favor gravitational instability in protostellar disks \citep{Rice2003} rather than fragmentation of proto-stellar clouds \citep{Bate2000} as the formation mechanism of brown dwarf-mass companions because stellar systems with larger difference in mass between primary and secondary stars are more difficult to form by the latter mechanism \citep{Bate2000}.

Also, there seems to be a possible paucity of lower-mass companions around 2.4$\sim$2.6--4 $M_{\odot}$ stars, in particular, a lack of planetary companions around 3--4 $M_{\odot}$ stars (region (b) in figure \ref{fig1}). Although it is basically difficult to detect planets around such "noisy" stars with large intrinsic radial velocity variability ($\sigma$ $\sim$ 20 m s$^{-1}$), planets with mass $\geq$ 2.6--3.3 $M_{\mathrm{J}}$ ($\geq$ 5.7--7.4 $M_{\mathrm{J}}$) and $a$ $=$ 0.6 AU ($a$ $=$ 3.0 AU) should be above the current detection limit (3 $\sigma$ $\sim$ 60 m s$^{-1}$). \citet{Kennedy2008} predicted that the frequency of giant planets has a peak near 3 $M_{\odot}$ stars based on a core accretion scenario taking account of the movement of snow line along the evolution of accretion and the central stars. Moreover, if a formation mechanism works that invokes capturing of solid bodies migrating inward at the inner edge of the inactive magnetorotational instability-dead zone inside of the protoplanetary disk, gas giant planets could be formed efficiently at around 1 AU around intermediate-mass stars before the planetary disks deplete \citep{Kretke2009}. Increasing the number of known massive planetary companions around massive intermediate-mass stars by further radial velocity surveys would be of great interest to understand the formation mechanisms of giant planets around intermediate-mass stars.

\begin{figure}
  \includegraphics[height=.45\textheight]{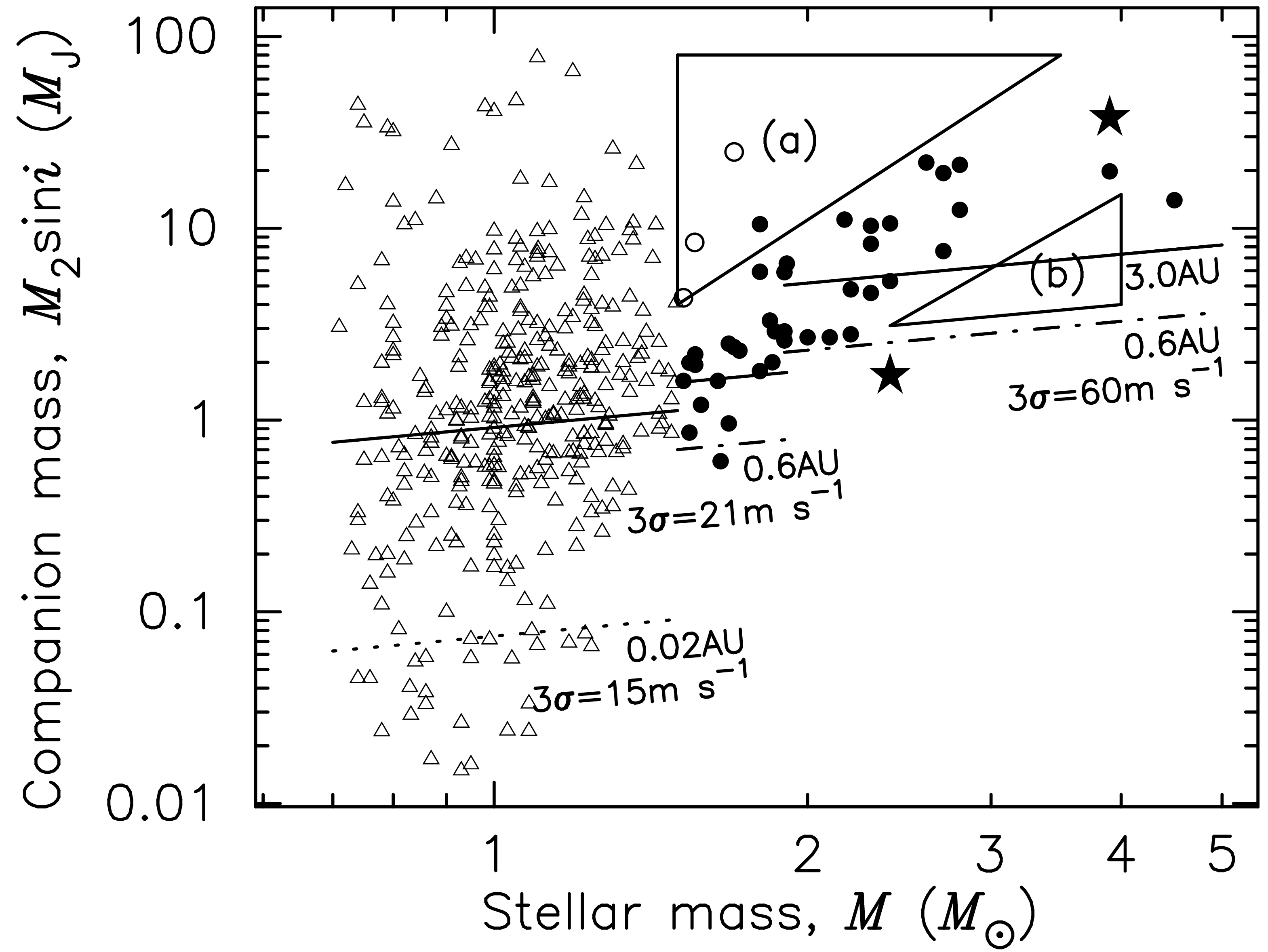}
  \caption{Primary star masses versus masses of substellar companions orbiting within 3 AU. \textit{Open triangles}, \textit{filled circles} and \textit{open circles} represent companions orbiting solar-mass stars, intermediate-mass evolved stars (subgiants and giants) and intermediate-mass dwarfs, respectively. \textit{Stars} represent the HD 119445 and the giant star. Solid lines indicate the detection limits for the mass of companions orbiting at 3 AU, corresponding to three times of typical radial velocity jitters $\sigma$ of 5 m s$^{-1}$ for solar-mass stars (0.7 $M_{\odot}$ $\leq$ $M$ $<$ 1.5 $M_{\odot}$), 7 m s$^{-1}$ for  intermediate-mass subgiants (1.5 $M_{\odot}$ $\leq$ $M$ $\leq$ 1.9 $M_{\odot}$) and 20 m s$^{-1}$ for  intermediate-mass giants (1.9 $M_{\odot}$ $<$ $M$ $\leq$ 5 $M_{\odot}$). \textit{Dotted} and \textit{dot-dashed lines} indicate the detection limits for companions at 0.02 AU and 0.6 AU in solar mass stars and intermediate-mass evolved stars, respectively. Two regions devoid of substellar companions are denoted by (a) and (b). }
  \label{fig3}
\end{figure}

\begin{theacknowledgments}
This research was supported as a Korea-Japan Joint Research Project under the Japan-Korea Basic Scientific Cooperation Program between Korea Science and Engineering Foundation (KOSEF) and Japan Society for the Promotion of Science (JSPS). 
\end{theacknowledgments}


\end{document}